\documentclass[sigconf]{acmart}

\AtBeginDocument{%
  }

\copyrightyear{2026}
\acmYear{2026}
\setcopyright{cc}
\setcctype{by}
\acmConference[RecSys '26]{20th ACM Conference on Recommender Systems}{September 27-October 02, 2026}{Minneapolis, MN, USA}
\acmBooktitle{20th ACM Conference on Recommender Systems (RecSys '26), September 27-October 02, 2026, Minneapolis, MN, USA}
\acmDOI{10.1145/3773078.3841297}
\acmISBN{979-8-4007-2284-4/2026/09}

\usepackage{pifont}
\usepackage{tikz}
\usetikzlibrary{arrows.meta,positioning,fit,backgrounds,calc}

\begin{document}

\title{Conversational Recommendation over Live E-Commerce Catalogues with Self-Refreshing Retrieval}

\author{Ante Kapetanović}
\affiliation{%
  \institution{Infobip}
  \city{Split}
  \country{Croatia}}
\email{ante.kapetanovic@infobip.com}

\author{Tomislav Đuričić}
\affiliation{%
  \institution{Infobip}
  \city{Zagreb}
  \country{Croatia}}
\email{tomislav.duricic@infobip.com}

\author{Dionizije Fa}
\affiliation{%
  \institution{Infobip}
  \city{Osijek}
  \country{Croatia}}
\email{dionizije.fa@infobip.com}

\author{Andro Merćep}
\affiliation{%
  \institution{Infobip}
  \city{Zagreb}
  \country{Croatia}}
\email{andro.mercep@infobip.com}

\author{Emanuel Lacić}
\affiliation{%
  \institution{Infobip}
  \city{Zagreb}
  \country{Croatia}}
\email{emanuel.lacic@infobip.com}

\renewcommand{\shortauthors}{Ante Kapetanovic, Tomislav Duricic, Dionizije Fa, Andro Mercep, and Emanuel Lacic}

\begin{abstract}
Conversational recommender systems based on large language models (LLMs) are usually evaluated on static, pre-indexed item collections, yet e-commerce catalogues change continuously as products are added or removed, repriced, and restocked.
We present a merchant-agnostic, multi-turn conversational shopping assistant that operates over such live catalogues.
Its central component is a self-refreshing retriever that ingests a merchant product feed, enriches the records, and synchronizes them into a vector index.
On each run, per-item hashes identify which products are new, changed, deleted, or unchanged, so only the delta is processed rather than rebuilding the whole catalogue.
A controller-based dialogue layer consumes this index, using an LLM only for intent classification and preference elicitation while retrieval, reranking, and diversity selection run as dedicated functions.
Our demonstration is a WhatsApp shopping assistant in which catalogue changes reach the recommendations after the next successful sync.
A live chatbot, documentation, and a recorded walkthrough are available at [\url{https://github.com/infobip/infobip-agentic-crs}].
\end{abstract}



\maketitle

\section{Introduction}
Most large language model (LLM)-based conversational recommender systems (CRSs) are evaluated over fixed benchmark collections~\cite{he2023zeroshot,jannach2021survey}, but in production the catalogue is a live object, continually updated.
Re-indexing the whole catalogue on every change is wasteful, yet letting the index drift degrades recommendations and surfaces out-of-stock or discontinued items.

Our work sits within LLM-era CRS~\cite{jannach2021survey,kolb2025tutorial}: retrieval-augmented CRSs ground recommendations in an item corpus~\cite{yang2024retrieval,kemper2024racr,lewis2020rag}, agentic designs give the LLM tools and control flow~\cite{huang2025recagent,yao2023react,schick2023toolformer}, and memory-enhanced systems enrich dialogue context~\cite{xi2024memocrs}, but deployed LLM-driven CRSs remain rare~\cite{kunstmann2026eventchat}.
Our emphasis is orthogonal to model quality.
We treat catalogue freshness---keeping the index consistent with a live assortment---as the engineering problem that makes such systems production-viable, complementing work on adapting LLM recommenders to refreshed indices~\cite{he2025reindex}.
The engine is merchant-agnostic, reusing an existing product feed.

\begin{figure}[t]
    \centering
    \includegraphics[width=0.4\linewidth]{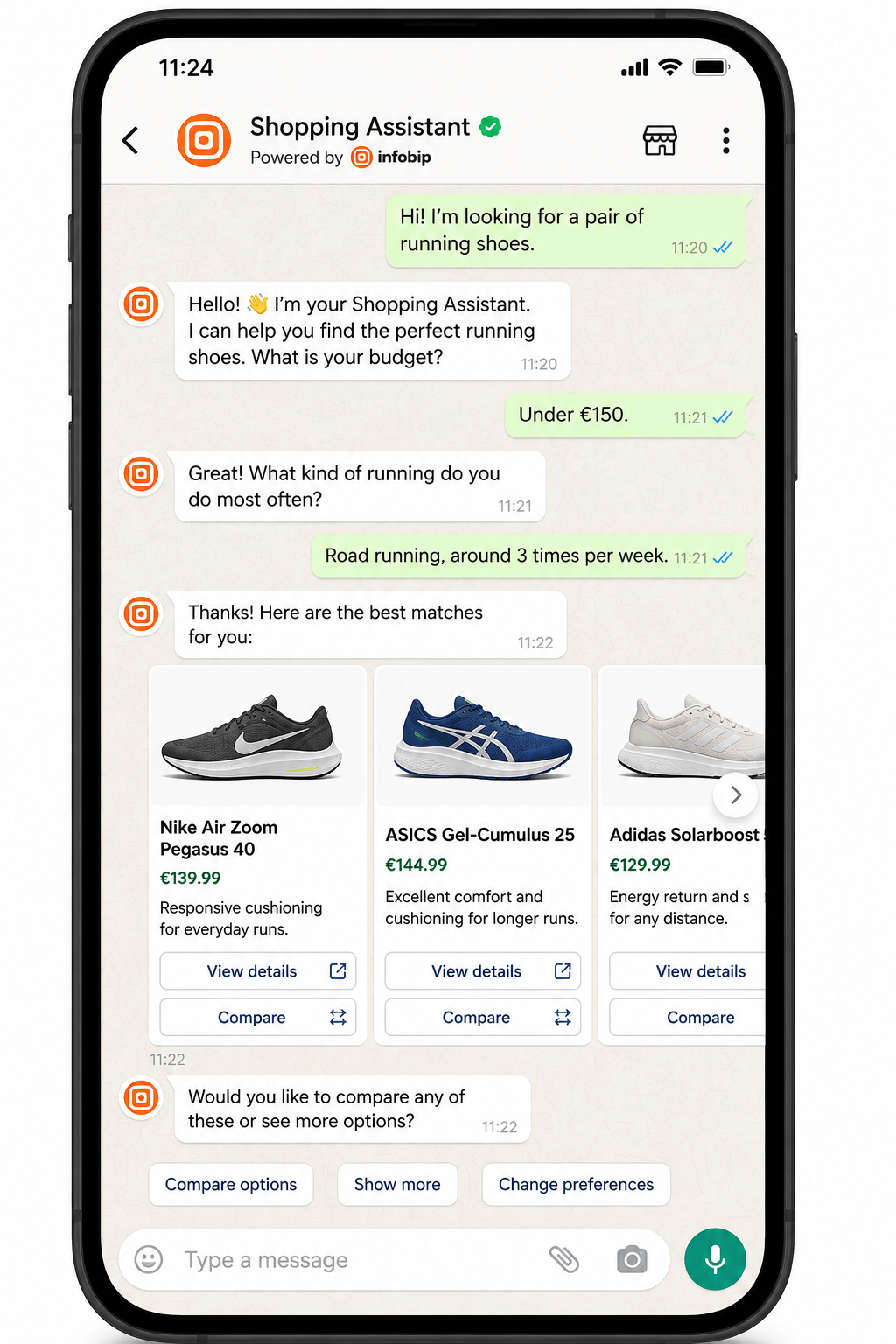}
    \caption{Example multi-turn WhatsApp interaction showing preference elicitation and product recommendation.}
    \Description{Screenshot of a multi-turn WhatsApp chat in which the assistant asks clarifying questions about the shopper's preferences and then replies with a short list of recommended products and links.}
    \label{fig:whatsapp-demo}
\end{figure}

We demonstrate a conversational shopping assistant built around one contribution: a self-refreshing retriever that re-embeds only new or semantically changed products, keeping synchronization proportional to the changed subset.
A dialogue pipeline consumes the refreshed index independently of the vector store, model provider, and channel, reaching WhatsApp through Infobip Answers.\footnote{\url{https://www.infobip.com/docs/answers}}

\section{System Overview}
The engine has three subsystems (Appendix~\ref{app:arch}): a catalogue pipeline that ingests and indexes products, a conversation pipeline that handles multi-turn dialogue, and a storage layer, written by the former and read by the latter, providing vector search, user profiles, and session state.
This shared storage decouples catalogue synchronization from dialogue, allowing each pipeline to run independently.
All generative, embedding, and reranking calls use a single proxy, making model choices configuration rather than code.
Our proof of concept uses ChromaDB through a swappable \texttt{VectorStore} interface.

\subsection{Self-Refreshing Retriever}
The retriever refreshes a searchable vector index from a merchant product feed.
Each manual or scheduled run compares the latest catalogue snapshot with the index and applies only the difference; it does not monitor the feed continuously.

\paragraph{Fetch and parse.} A run reuses a fresh cached feed; otherwise, it streams XML to disk and retries failed downloads with exponential backoff, deleting an incomplete file after the final failure. An incremental parser bounds memory use, strips HTML, normalises prices, and converts availability to a boolean.

\paragraph{IDs, hashes, and embeddings.} A stable product ID links snapshots and drives exact updates and deletions, but cannot answer natural-language queries.
A full hash detects any feed-field change; a semantic hash over name, description, brand, and category identifies changes requiring re-embedding.
The resulting vectors make products retrievable; the generative LLM is only an enrichment fallback.

\paragraph{Change classes.} Comparing IDs and hashes yields five disjoint classes.
New and semantically changed records are enriched, embedded, and upserted; enrichment resolves category paths and extracts attributes by rule, with generative fallback.
Metadata-only changes, such as price or stock, retain the vector while updating the record and filters.
Deleted records are removed, and unchanged records are skipped.

\subsection{Conversational Pipeline}
The conversation pipeline follows an orchestrator-as-controller pattern~\cite{yao2023react,schick2023toolformer,huang2025recagent}: a generative model classifies messages into eight intents, composes replies, and uses an elicitor sub-agent to ask one to three clarifying questions when preferences are vague~\cite{shimazu2001expertclerk,sun2018crs}.
Recommendation uses content-based semantic retrieval: query and product text share one embedding space, metadata filters restrict candidates, an optional non-generative model reranks them~\cite{yang2024retrieval,kemper2024racr}, and a greedy selector adds brand and category variety.
This generation-free path keeps cost predictable, although embedding and reranking still call external models~\cite{kolb2025tutorial}; the pipeline detects the user's language, retrieves in English, and replies in that language.

\section{Demonstration}
Users access the live demonstration on any smartphone through WhatsApp, with no application to install (Figure~\ref{fig:whatsapp-demo}).
Sessions may be anonymous or personalised from prior purchases; the assistant elicits preferences, searches the live catalogue, and returns diverse in-stock products with links, and each successful sync exposes catalogue changes.

\begin{table}[t]
\centering
\small
\caption{Incremental synchronization of an anonymized 500-record catalogue (medians over three or five runs; full rebuild: 2.914\,s).
\ding{51}/\ding{55} indicate enrichment (E), embedding (Emb), metadata update (M), and deletion (D).}
\Description{Rows show five catalogue change scenarios. Columns indicate whether each scenario triggers enrichment, embedding, metadata update, or deletion, followed by runtime and percentage of full-rebuild time.}
\label{tab:sync}
\begin{tabular}{@{}lcccccc@{}}
\toprule
Change & E & Emb & M & D & Time (s) & Full (\%) \\
\midrule
None                 & \ding{55} & \ding{55} & \ding{55} & \ding{55} & 0.053 & 1.8 \\
Add product          & \ding{51} & \ding{51} & \ding{55} & \ding{55} & 0.321 & 11.0 \\
Price/stock          & \ding{55} & \ding{55} & \ding{51} & \ding{55} & 0.072 & 2.5 \\
Description/category & \ding{51} & \ding{51} & \ding{55} & \ding{55} & 0.357 & 12.3 \\
Delete product       & \ding{55} & \ding{55} & \ding{55} & \ding{51} & 0.062 & 2.1 \\
\bottomrule
\end{tabular}
\end{table}

Table~\ref{tab:sync} confirms the intended change classification: price or stock changes update only metadata, description or category changes re-enrich and re-embed the record, and all ID, hash, feed-field, and embedding-call checks passed.

The prototype accepts Google Merchant Center Atom feeds directly; other catalogue sources or messaging channels require adapters to the product schema or engine API, while filtering, batching, consistency, and operations remain backend concerns behind the \texttt{VectorStore} interface.
Synchronization rejects parser failures, invalid or duplicate IDs, and incompatible index versions, and rejects empty or incomplete snapshots to prevent inferred mass deletion; it prepares enrichment and embeddings before writing, then applies upserts and metadata updates before deletions.
Writes are non-transactional, so production requires truncated-feed detection, staging or rollback, and abnormal-delta monitoring; we make no tens-of-millions-scale claim.
Documentation, engine excerpts, a demo video, and a runnable synthetic sync are public; the full engine, catalogue, and Infobip integration remain private.

\section{Concluding Remarks}
We presented a conversational shopping assistant whose self-refreshing retriever keeps a vector index consistent with a live catalogue, processing only new, changed, or removed products on each sync; a measured case study confirms the classification and operation counts behave as intended, and the assistant is deployed as a WhatsApp demo.
The evaluation covers synchronization not ranking quality; an offline relevance study and a live user study of recommendation quality and cost remain future work, alongside freshness-aware retrieval and hybrid ranking.

\begin{acks}
This research was supported in part by the project Infobip Global Communication Platform (PK.1.1.07.0001), part of the Important Project of Common European Interest on Next Generation Cloud Infrastructure and Services (IPCEI-CIS) consortium.

Generative AI tools were used in a supporting capacity: Claude Code (Opus 4.7) for code implementation and data analysis, Chat-GPT 5.5 for \LaTeX{} editing and grammar checking. All AI-assisted outputs were reviewed and approved by the authors, who take full responsibility for the content of this publication.
\end{acks}

\appendix
\begin{figure*}[t!]
\centering
\definecolor{cRetr}{RGB}{37,99,235}
\definecolor{cStore}{RGB}{22,163,74}
\definecolor{cConv}{RGB}{124,58,237}
\resizebox{\textwidth}{!}{%
\begin{tikzpicture}[
  font=\footnotesize,
  node distance=6mm and 7mm,
  box/.style={draw, semithick, rounded corners=2pt, align=center,
              minimum height=9mm, inner sep=3pt, text width=19mm, fill=white},
  retr/.style={box, draw=cRetr, fill=cRetr!8},
  stor/.style={box, draw=cStore, fill=cStore!10},
  conv/.style={box, draw=cConv, fill=cConv!8},
  arr/.style={-{Latex[length=2mm]}, semithick, black!65},
  biarr/.style={{Latex[length=2mm]}-{Latex[length=2mm]}, semithick, black!65},
  darr/.style={arr, dashed},
  elbl/.style={font=\scriptsize, text=black!55, fill=white, inner sep=1pt},
]
\node[retr] (feed)   {Product feed\\(XML)};
\node[retr, right=of feed]     (fetch)    {Fetch\\$+$ cache};
\node[retr, right=of fetch]    (parse)    {Parse\\(streaming)};
\node[retr, right=of parse]    (classify) {Classify by ID,\\full/semantic\\hash};
\node[retr, right=of classify] (enrich)   {Enrich\\taxonomy, attrs};
\node[retr, right=of enrich]   (embed)    {Embed};
\foreach \a/\b in {feed/fetch,fetch/parse,parse/classify}
  \draw[arr] (\a) -- (\b);
\draw[arr] (classify) -- (enrich) node[elbl, pos=0.5, above] {new / sem.};
\draw[arr] (enrich) -- (embed);
\node[cRetr, font=\bfseries, anchor=south west] at (feed.north west)
  {Self-refreshing retriever};
\node[draw=black!45, dashed, rounded corners, fit=(enrich)(embed), inner sep=5pt,
      label={[font=\itshape, text=black!55]above:generative / embedding proxy}] {};

\node[stor, text width=58mm] (store) at ($(feed)!0.5!(embed)+(0,-1.9)$)
  {Storage layer: vector store (\texttt{VectorStore} interface)\\$+$ user profiles $+$ session state};
\draw[arr] (embed.south) |- (store.north east)
  node[elbl, pos=0.72] {upsert $\Delta$};
\draw[arr] (classify.south) -- (classify.south |- store.north)
  node[elbl, pos=0.5, right] {metadata update / delete};
\node[elbl] at ($(parse)!0.5!(classify)+(0,-0.95)$) {unchanged: skip};

\node[conv] (orch) at ($(store.south)+(0,-1.2)$) {Orchestrator\\intent $+$ dispatch};
\node[conv, left=16mm of orch] (user) {User via\\WhatsApp};
\node[conv, text width=32mm, right=16mm of orch] (rec)
  {Recommend\\search $\to$ rerank $\to$ diversity};
\node[conv, below=5mm of orch] (elic) {Elicitor (LLM)};
\node[cConv, font=\bfseries, anchor=south west] at (user.north west)
  {Conversation pipeline};
\draw[arr]  (user) -- (orch);
\draw[arr]  (orch) -- (rec);
\draw[biarr] (orch) -- (elic);
\draw[darr] (store.south) -- (orch.north) node[elbl, pos=0.5] {profiles};
\draw[arr]  (rec.north) |- (store.south east) node[elbl, pos=0.72] {search};
\end{tikzpicture}}
\caption{Engine architecture.
\textbf{Top:} the \textcolor{cRetr}{self-refreshing retriever} turns a merchant product feed into a vector index.
It fetches and parses the feed, then classifies each item by its stable ID and its full and semantic content hashes.
New and semantically changed items are enriched, embedded, and upserted.
Metadata-only changes update the stored record while keeping its vector.
Deleted items are removed, and unchanged items are skipped.
\textbf{Bottom:} the \textcolor{cConv}{conversation pipeline} serves shoppers over WhatsApp.
The orchestrator classifies intent and dispatches either to the elicitor or to the recommend path (search $\to$ rerank $\to$ diversity), calling a generative model only for intent and elicitation.
\textbf{Center:} a \textcolor{cStore}{shared storage layer} (vector store behind a \texttt{VectorStore} interface, user profiles, sessions) is written by the retriever and read by the conversation pipeline.}
\Description{Block diagram of the engine's three subsystems. The top band shows the self-refreshing retriever as a pipeline: product feed, fetch and cache, streaming parse, and a classify stage that compares each item's stable ID and its full and semantic hashes. New and semantically changed items continue through enrich and embed stages and are upserted into storage. Metadata-only changes update the stored record and keep its vector. Deleted items are removed and unchanged items are skipped. The center band is a shared storage layer holding the vector store, user profiles, and session state, written by the retriever and read by the conversation pipeline. The bottom band shows the conversation pipeline: a user on WhatsApp talks to an orchestrator that dispatches to an elicitor or to the search, rerank, and diversity recommend path. Enrichment fallback, embedding, and reranking calls route through a single proxy.}
\label{fig:arch}
\end{figure*}
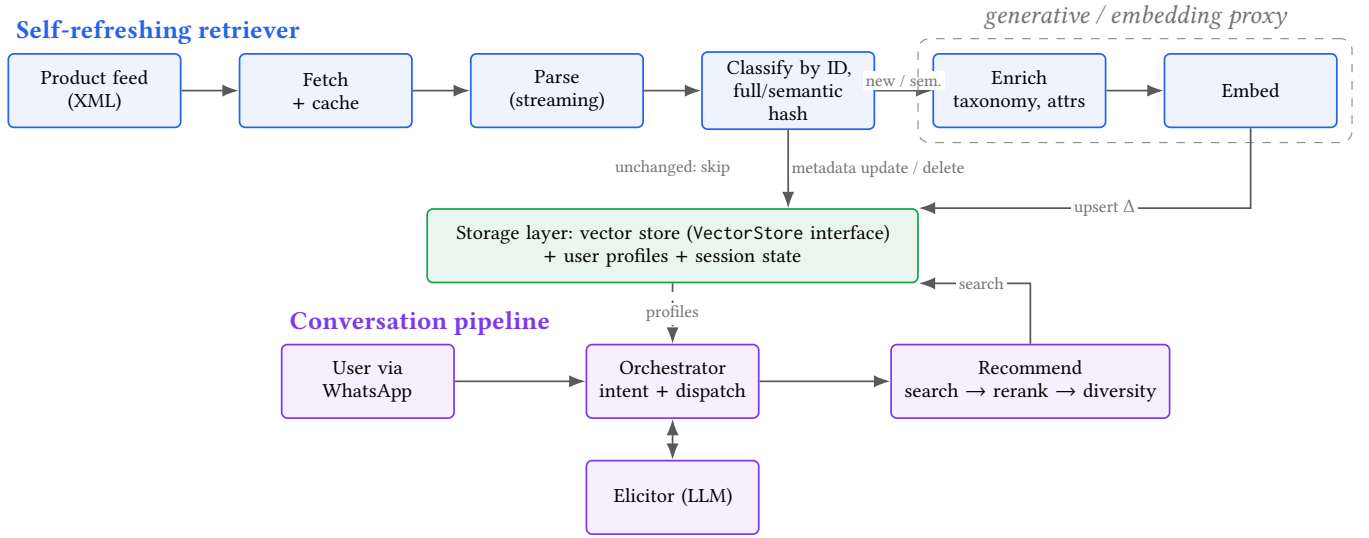

\section{Engine Architecture}
\label{app:arch}
Figure~\ref{fig:arch} details the engine's three subsystems and how they share state.
The self-refreshing retriever (top) turns a merchant feed into a vector index, classifying each item by its stable ID and its full and semantic hashes so that only new or semantically changed items are enriched and embedded.
The conversation pipeline (bottom) serves WhatsApp shoppers through an orchestrator that dispatches to the elicitor or to the recommend path (search $\to$ rerank $\to$ diversity).
Both meet at a shared storage layer, holding vectors, profiles, and sessions (retriever writes, conversation pipeline reads).

\bibliographystyle{ACM-Reference-Format}
\bibliography{references}

\end{document}